## Particle resonance in the (1+1)-dimensional Dirac equation with kink-like vector potential and delta interaction

M. Eshghi[*,1], H. Mehraban[2] , Sameer M. Ikhdair [3]

[1] *Young Researchers and Elite club, Central Tehran Branch, Islamic Azad University, Tehran, Iran*

[2] *Faculty of Physics, Semnan University, Semnan, Iran*

[3]*Department of Physics, Near East University, Nicosia, Northern Cyprus, Turkey*

*and*

[3] *Department of Physics, Faculty of Science, An-Najah National University, Nablus, West Bank, Palestine*

## Abstract

The relativistic problem of spin-$1/2$ fermions subject to vector hyperbolic (kink-like) potential $(\sim \tanh kx)$ is investigated by using the parametric Nikiforov-Uvarov method. The energy eigenvalue equation and the corresponding normalized wave functions are obtained in terms of the Jacobi polynomials for $x>0$ and $x<0$ cases.

### 1. Introduction

The search for the exact solutions of relativistic wave equations under the direct coupling of various vector and scalar potentials has been an important research area ever since the birth of quantum mechanics, and has significantly enriched our knowledge of the atomic and sub-atomic systems. In fact, the success of quantum mechanics in the description of the atomic and sub-micro world is very impressive and overwhelming. Supplementing this theory with special relativity created one of the most accurate physical theories in recent history. An example is quantum

[1] *Corresponding author*: *Tel*.:+982177104932; *fax*: +982177104938*; Email***:** eshgi54@gmail.com

.

electrodynamics: the theory that describes the interaction of charged particles with the electromagnetic radiation at high speeds or strong coupling. The Dirac equation is the most frequently used wave equation for the description of particle dynamics in relativistic quantum mechanics and in many fields of physics and chemistry. For this reason, it has been studied and used extensively in relativistic heavy ion collisions, heavy ion spectroscopy and more recently, in laser–matter interaction (for a review, see [1] and references therein) and condensed matter physics [2]. However, solving this equation is still a very challenging problem even if it has been derived more than 80 years ago and has been utilized profusely. This equation is very useful to investigate the relativistic effects as well [3]. In the relativistic treatment of nuclear phenomena the Dirac equation is used to describe the behavior of nucleus in nuclei. In fact, when the particle is under strong field, especially for a strong coupling system, relativistic effect could become important. In the strong coupling case, relativistic effects have been rarely discussed, primarily due to difficulties involved in solving analytically the Klein-Gordon equation or the Dirac equation. The Dirac equation, which describes the motion of a spin-1/2 particle, has been used in solving many problems of nuclear and high-energy physics. Several model potentials have been introduced recently to explore the relativistic energy spectra and wave function behaviors (for example, see [4-13]). The hyperbolic potential [14, 15] is given by

$$V(x) = \Lambda \tanh(kx) \tag{1}$$

where $\{\Lambda, k\} \subset \mathbb{R}$ are empirical constants. The potential (1) is also called kink-like potential. For this Kink-like potential, there exists no bound states in a non-relativistic Schrödinger quantum theory because it gives rise to a ubiquitous repulsive potential. However, bound states of this Kink-like potential exist in the Dirac theory and the Klein-Gordon (KG) equation. This potential is a kind of special Rosen-Morse II potential and asymptotically constant for large values of $x$ and has the linear potential as a limit for small values of $k$. The potential can be used to describe the nuclei interactions or the quark physics. By using the algebraic method, Wen-Jia Tian has solved the Dirac equation for *s*-wave and the KG equation with the kink-like potential [16]. de Castro and Hott [17] investigated the relativistic problem of trapping neutral fermions subject to a pseudoscalar kink-like potential. The bound states of this kink-like potential exist in (1+1)-dimensional Dirac equation with pseudoscalar potential coupling. de Castro [18] also investigated the intrinsically relativistic problem of spinless particles in (1+1)-dimensional KG equation subject to a general mixing of vector and scalar kink-like potentials $(\sim \tanh \gamma x)$ coupling in two-dimensional space-time. The problem was mapped into the exactly solvable Sturn-Liouville problem with the Rosen-Morse potential and exact bounded solutions for particles and antiparticles were found. The behavior of the spectrum was discussed and the apparent paradox concerning the uncertainty was solved by recurring to the concept of effective Compton wavelength. Jia and co-workers [19] studied the bounded solutions of the 1+1 dimensional Dirac equation and KG equation with a PT-symmetric version of the kink-like vector potential in two-dimensional space-time by using the basic concepts of the Supersymmetric, WKB formalism and the function analysis method..They obtained the bound-state energy levels and two-spinor components. The PT-symmetric kink-like potential is not Hermitian and absent bound states in the context of non-relativistic Schrödinger equation but it possesses two sets of real discrete relativistic energy spectra in the context of the Dirac theory. Jia and Souza-Dutra [20], solved position-dependent mass Dirac equation with the vector potential coupling scheme in 1+1 dimensions. They presented three PT-symmetric potential harmonic oscillator-like potential, PT-symmetric with the form of a linear potential plus an inversely linear potential and PT symmetric kink-like potential.

Recently Jia et al. [21] studied the exact solutions of the KG equation with position-dependent mass for mixed vector and scalar kink-like potentials. Villalba and González-Díaz [22] showed that the energy spectrumof the one-dimensional Dirac equation in the presence of an attractive vectorial delta potential exhibits a resonant behavior when one includes an asymptotically spatially vanishing weak electric field associated with a hyperbolic tangent potential. The resonant behavior depends on the strength of electric field. They also derived an approximate expression for the value of the resonances and compared the obtained results for the hyperbolic potential with those obtained for a linear perturbative potential.

Our aim in this paper is to solve the Dirac equation for the above Kink-like potential. Thus, we obtain the energy eigenvalues equation and the corresponding spinor wave functions by using the parametric generalization of the Nikiforov-Uvarov (NU) method.

The paper is structured as follows: In section 2, we introduce the outlines of the parametric Nikiforov-Uvarov method. In section 3, we solve the (1+1)-dimensional Dirac equation in the presence of attractive $\delta$.potential and a kink-like vector potential. We also derive approximate analytic expressions for the energy eigenvalues and wave functions for $x>0$ and $x<0$ cases. In Section 4, we end with our conclusions.

## 2. Parametric NU method

The NU method is used to solve second order differential equations with an appropriate coordinate transformation $s=s(r)$ [23]

$$\left[\frac{d^2}{ds^2}+\frac{\tilde{\tau}(s)}{\sigma(s)}\frac{d}{ds}+\frac{\tilde{\sigma}(s)}{\sigma^2(s)}\right]\psi_n(s)=0, \tag{2}$$

where $\sigma(s)$ and $\tilde{\sigma}(s)$ are polynomials with at most of second degree, and $\tilde{\tau}(s)$ is a first-degree polynomial. To make the application of the NU method simpler and direct without need to check the validity of solution. We present a shortcut for the method. So, at first we write the general form of the Schrödinger-like equation (2) in a more general form applicable to any potential as follows [24-26]

$$\left[\frac{d^2}{ds^2}+\frac{c_1-c_2s}{s(1-c_3s)}\frac{d}{ds}+\frac{1}{s^2(1-c_3s)^2}\left(-As^2+Bs-C\right)\right]\psi_n(s)=0, \tag{3}$$

satisfying the wave functions

$$\psi_n(s)=\phi(s)Y_n(s). \tag{4}$$

Comparing (3) with its counterpart (2), we can obtain the following identifications:

$$\tilde{\tau}(s)=c_1-c_2s,\quad \sigma(s)=s(1-c_3s),\quad \tilde{\sigma}(s)=-As^2+Bs-C, \tag{5}$$

(1) For the given choice of root $k_-$ and the function $\pi(s)$:

$$k_-=-(c_7+2c_3c_8)-2\sqrt{c_8c_9},\quad \pi(s)=c_4+\sqrt{c_8}-\left(\sqrt{c_9}+c_3\sqrt{c_8}-c_5\right)s, \tag{6}$$

we follow the NU method to obtain the eigenfunctions Eq. (3) as follows [25,26]

$$\begin{aligned}
&\rho(s)=s^{c_{10}}(1-c_3s)^{c_{11}},\\
&\phi(s)=s^{c_{12}}(1-c_3s)^{c_{13}},\ c_{12}>0,\ c_{13}>0,\\
&Y_n(s)=\frac{1}{\rho(s)}\frac{d^n}{ds^n}\left[\sigma^n(s)\rho(s)\right]\sim P_n^{(c_{10},c_{11})}(1-2c_3s),\ c_{10}>-1,\ c_{11}>-1,
\end{aligned} \tag{7}$$

$$\psi_n(s) = N_n s^{c_{12}} (1-c_3 s)^{c_{13}} P_n^{(c_{10},c_{11})}(1-2c_3 s).$$

where $P_n^{(\mu,\nu)}(x),\ (\mu > -1,\ \nu > -1 \text{ and } x \in [-1,1])$ is the Jacobi polynomial and $\rho(s)$ is the weight function. Further, the energy equation reads

$$nc_2 - (2n+1)c_5 + (2n+1)\left(\sqrt{c_9} + c_3\sqrt{c_8}\right) + n(n-1)c_3 + c_7 + 2c_3c_8 + 2\sqrt{c_8c_9} = 0, \qquad (8)$$

and with the parametric constants [25,26]

$$\begin{aligned} c_4 &= \frac{1}{2}(1-c_1), & c_5 &= \frac{1}{2}(c_2 - 2c_3), \\ c_6 &= c_5^2 + p_2; & c_7 &= 2c_4c_5 - p_1, \\ c_8 &= c_4^2 + p_0, & c_9 &= c_3(c_7 + c_3c_8) + c_6, \\ c_{10} &= 2\sqrt{c_8} > -1, & c_{11} &= \frac{2}{c_3}\sqrt{c_9} > -1,\ c_3 \neq 0, \\ c_{12} &= c_4 + \sqrt{c_8} > 0, & & \\ & & c_{13} &= -c_4 + \frac{1}{c_3}(\sqrt{c_9} - c_5) > 0,\ c_3 \neq 0, \end{aligned} \qquad (9)$$

where $c_{12} > 0,\ c_{13} > 0$ and $s \in [0, 1/c_3],\ c_3 \neq 0.$

In a rather more special case where $c_3 = 0$, the wave functions (6) can be expressed instead as

$$\begin{aligned} &\lim_{c_3 \to 0} P_n^{(c_{10},c_{11})}(1-2c_3 s) = L_n^{c_{10}}\left(2\sqrt{c_9}\, s\right),\ \lim_{c_3 \to 0}(1-c_3 s)^{c_{13}} = e^{-(\sqrt{c_9}-c_5)s}, \\ &\psi(s) = N s^{c_{12}} e^{-(\sqrt{c_9}-c_5)s} L_n^{c_{10}}(2\sqrt{c_9}\, s) \end{aligned} \qquad (10)$$

where $L_n^{(\lambda)}(\tau x)$ is the Laguerre polynomial.

(2) For the given root $k_+$ and the function $\pi(s)$:

$$k_+ = -(c_7 + 2c_3c_8) + 2\sqrt{c_8c_9},\ \ \pi(s) = c_4 - \sqrt{c_8} - \left(\sqrt{c_9} - c_3\sqrt{c_8} - c_5\right)s, \qquad (11)$$

we follow the NU method [23] to obtain the energy equation

$$nc_2 - (2n+1)c_5 + (2n+1)\left(\sqrt{c_9} - c_3\sqrt{c_8}\right) + n(n-1)c_3 + c_7 + 2c_3c_8 - 2\sqrt{c_8c_9} = 0, \qquad (12)$$

and also the wave functions

$$\begin{aligned} &\rho(s) = s^{\tilde{c}_{10}}(1-c_3 s)^{\tilde{c}_{11}},\ \ \phi(s) = s^{\tilde{c}_{12}}(1-c_3 s)^{\tilde{c}_{13}},\ \tilde{c}_{12} > 0,\ \tilde{c}_{13} > 0, \\ &y_n(s) = P_n^{(\tilde{c}_{10},\tilde{c}_{11})}(1-2\tilde{c}_3 s),\ \tilde{c}_{10} > -1,\ \tilde{c}_{11} > -1, \end{aligned} \qquad (13)$$

$$\psi_n(s) = N_{n\kappa} s^{\tilde{c}_{12}} (1-c_3 s)^{\tilde{c}_{13}} P_n^{(\tilde{c}_{10},\tilde{c}_{11})}(1-2c_3 s),$$

with the following changes in parametric constants:

$$\tilde{c}_{10} = -2\sqrt{c_8}, \qquad \tilde{c}_{11} = \frac{2}{c_3}\sqrt{c_9},\ c_3 \neq 0,$$

$$\tilde{c}_{12} = c_4 - \sqrt{c_8} > 0. \tag{14}$$

$$\tilde{c}_{13} = -c_4 + \frac{1}{c_3}(\sqrt{c_9} - c_5) > 0,\ c_3 \neq 0.$$

### 3. The (1+1)-Dimensional Dirac Equation

The Dirac equation for fermionic massive spin-$1/2$ particles moving in a hyperbolic vector potential $V(r)$, expressed in natural units ($\hbar = c = 1$) takes the form [27]

$$\left[ i\gamma^{\mu}\left( \frac{\partial}{\partial x^{\mu}} - ieA_{\mu} \right) - m \right]\psi(\vec{r}) = 0, \tag{15}$$

where $A_{\mu}$ is the vector potential that, in our case, takes the form $eA^{\mu} = \left(-g\,\delta(x) + V(x)\right)\delta_0^{\mu}$, where $e$ is the charge and $m$ is the mass of the fermionic particle.

$\gamma^{\mu}$ $(\mu = 0,1,2,3)$ are gamma matrices and satisfy the commutation relation $\left\{\gamma^{\mu},\gamma^{\nu}\right\} = 2g^{\mu\nu}$ with $g^{\mu\nu} = \mathrm{diag}(1,-1)$ [28]. The Dirac matrices

$$\gamma^0 = \begin{pmatrix} 0 & I \\ I & 0 \end{pmatrix}, \qquad \gamma^i = \begin{pmatrix} 0 & -\sigma^i \\ \sigma^i & 0 \end{pmatrix}, \tag{16}$$

where $I$ is the $2\times 2$ unit matrix, and $\sigma^i$ are the three $2\times 2$ Hermitian Pauli matrices. In the absence of vector potential and setting $V(x) = eA_0(x)$, the one-dimensional Dirac equation for a stationary state $\Psi(x,t) = e^{-i\varepsilon t}\Psi(x)$ becomes [5]

$$\left\{ -i\frac{d}{dx}\begin{pmatrix} 0 & -1 \\ 1 & 0 \end{pmatrix} + \left(V(x) - E\right)\begin{pmatrix} 0 & 1 \\ 1 & 0 \end{pmatrix} + m\begin{pmatrix} 1 & 0 \\ 0 & 1 \end{pmatrix} \right\}\begin{pmatrix} \varphi(x) \\ \theta(x) \end{pmatrix} = 0, \tag{17}$$

where $E$ is the energy eigenvalue, $\varphi(x)$ and $\theta(x)$ are the upper and the lower components of the spinor wave function $\Psi(x)$, respectively. Equation (16) can be decomposed into the following two coupled differential equations [29]:

$$-i\frac{d\theta(x)}{dx}+\left(E-V(x)\right)\theta(x)-m\varphi(x)=0, \tag{18a}$$

$$i\frac{d\varphi(x)}{dx}+\left(E-V(x)\right)\varphi(x)-m\theta(x)=0, \tag{18b}$$

which is more tractable in the search of exact solutions. Eliminating the lower spinor component form Eqs. (18a) and (18b), we obtain a second order differential equation, which contains first order derivatives,

$$\frac{d^2\varphi(x)}{dx^2}+\left\{i\frac{dV(x)}{dx}+\left(V(x)-E\right)^2-m^2\right\}\varphi(x)=0. \tag{19}$$

### 3.1. Kink-like potential and a delta interaction

Now, applying the idea developed by Titchmarsch [30] and Barut [31], we obtain the energy spectrum of the (1+1)-dimensional Dirac equation in the presence of an attractive vector point interaction potential represented by $eV(x)=-g\delta(x),$ and an asymptotically vanishing electric field associated with the kink-like potential (1). The potential (1) is asymptotically constant for large values of $x$, i.e., $V(x)=\Lambda\lim_{x\to\infty}\frac{exp(kx)-exp(-kx)}{exp(kx)+exp(-kx)}\to\Lambda\sim\text{constant}$ and has the linear potential as a limit for small values of $k$, i.e., $V(x)=\Lambda\lim_{k\to 0}\frac{exp(kx)-exp(-kx)}{exp(kx)+exp(-kx)}\to\Lambda kx\sim x.$ This behavior of the kink-like potential is obvious in Figure 1. After substituting the potential (1) into Eq. (19), we obtain the second order differential equation

$$\frac{d^2\varphi(x)}{dx^2}+\left[ik\Lambda\sec h^2(kx)+\left(\Lambda\tanh(kx)-E\right)^2-m^2\right]\varphi(x)=0. \tag{20}$$

#### 3.1.1. The case of $x>0$

To obtain solution of Eq. (20) for $x>0$, exhibiting a damping asymptotic behavior as $x\to+\infty,$ we use the change of variables: $s=-e^{-2kx}$, to rewrite it as follows:

$$\left\{\frac{d^2}{ds^2}+\frac{1-s}{s(1-s)}\frac{d}{ds}+\frac{1}{s^2(1-s)^2}\times\right.$$

$$\left\{\left(\frac{-\left[m^2-(E+\Lambda)^2\right]s^2+(-4ik\Lambda+2\Lambda^2-2E^2+2m^2)s-\left(m^2-(E-\Lambda)^2\right)}{4k^2}\right)\right\}\varphi(s)=0, \qquad (21)$$

where $\varphi(s)$ satisfies the second order differential equation. Comparing Eq. (21) and Eq. (3), we can easily obtain the coefficients $c_i$ $(i=1,2,3)$ and analytical expressions $A$, $B$ and $C$ as follows

$$c_1=1, \qquad A=\frac{m^2-(E+\Lambda)^2}{4k^2},$$

$$c_2=1, \qquad B=\frac{(-4ik\Lambda+2\Lambda^2-2E^2+2m^2)}{4k^2},$$

$$c_3=1, \qquad C=\frac{m^2-(E-\Lambda)^2}{4k^2}. \qquad (22)$$

The specific values of coefficients $c_i$ $(i=4,5,...,13)$ are found from Eq. (9) and displayed in Table 1. The energy eigenvalues equation vcan be obtained via Eq. (8) as

$$\sqrt{\frac{k^2+4(ik\Lambda-\Lambda^2)}{k^2}}=\sqrt{\frac{m^2-(E+\Lambda)^2}{k^2}}-\sqrt{\frac{m^2-(E-\Lambda)^2}{k^2}}-(2n+1). \qquad (23)$$

Note that the resonant energies are obtained via Eq. (23). These approximations are valid when $\Lambda$ and $k$ are small compared to $m$. For large values of $k$, the kink-like potential approaches deeply which is not sink the delta bound state energies. These results are in agreement with [22] although the techniques used are different. However, the NU method provide simple and powerful results in closed form.

Since the parameters $\Lambda$ and $k$ are taken small in potential (1), we can approximate $(\Lambda\tanh kx-E)^2$ in Eq. (20) by $-2\Lambda kEx+E^2$ we can obtain an approximate differential equation for $x>0$. The solution of the resulting equation provides a very close solution to the exact one obtained via the above energy eigenvalues equation . This solution of the approximate kink-like solutions was studied in Ref. [22] in terms of the Airy function and in terms of the hypergeometric function for kink-like potential.

Now we need to find the corresponding wave functions by referring to Table.1 and Eq.(7), we find the necessary functions useful in calculating the wave function as

$$\rho(s)=s^{\frac{\sqrt{m^2-(E-\Lambda)^2}}{k}}(1-s)^{\frac{\sqrt{k^2+4(ik\Lambda-\Lambda^2)}}{k}}, \qquad (24)$$

$$\phi(s)=s^{\frac{\sqrt{m^2-(E-\Lambda)^2}}{2k}}(1-s)^{\frac{1}{2}\left(1+\frac{\sqrt{k^2+4\left(ik\Lambda-\Lambda^2\right)}}{k}\right)}, \tag{25}$$

$$y_n(s)\sim P_n^{\left(\sqrt{m^2-(E-\Lambda)^2}/k,\sqrt{k^2+4\left(ik\Lambda-\Lambda^2\right)}/k\right)}(1-2s), \tag{26}$$

By using $\varphi(s)=\phi(s)Y_n(s)$, we obtain the lower spinor radial wave functions

$$\begin{aligned}\varphi(x)&=Be^{-\sqrt{m^2-(E-\Lambda)^2}x}\left(1+e^{-2kx}\right)^{\left(1+\sqrt{k^2+4\left(ik\Lambda-\Lambda^2\right)}/k\right)/2}\\&\times P_n^{\left(\sqrt{m^2-(E-\Lambda)^2}/k,\sqrt{k^2+4\left(ik\Lambda-\Lambda^2\right)}/k\right)}\left(1+2e^{-2kx}\right),\end{aligned} \tag{27}$$

where $\sqrt{k^2+4\left(ik\Lambda-\Lambda^2\right)}/k$ is given in Eq. (23) and $B_n$ is the normalized constant. The above wave function satisfies the asymptotic conditions at $x=0$ and $x\to\infty$, i.e., $\varphi(x=0)$ is finite. And $\varphi(x\to\infty)\sim 0$. Now using the following definition of the Jacobi polynomial [32]

$$P_n^{(a,b)}(s)=\frac{\Gamma(n+a+1)}{n!\Gamma(a+1)}\,{}_2F_1\left(-n,a+b+n+1;a+1;\frac{1-s}{2}\right). \tag{28}$$

then Eq. (27) can be rewritten as

$$\begin{aligned}\varphi(x)&=B_ne^{-\sqrt{m^2-(E-\Lambda)^2}x}\left(1+e^{-2kx}\right)^{\left(1+\sqrt{k^2+4\left(ik\Lambda-\Lambda^2\right)}/k\right)/2}\\&\times{}_2F_1\left(-n,\sqrt{m^2-(E-\Lambda)^2}/k+\sqrt{k^2+4\left(ik\Lambda-\Lambda^2\right)}/k+n+1;\sqrt{m^2-(E-\Lambda)^2}/k+1;-e^{-2kx}\right).\end{aligned} \tag{29}$$

where $B_n=B\dfrac{\Gamma\left(n+\sqrt{m^2-(E-\Lambda)^2}/k+1\right)}{n!\Gamma\left(\sqrt{m^2-(E-\Lambda)^2}/k+1\right)}$.

Following Eq. (41) of Ref. [33] and using the differential and recursion properties of the Jacobi polynomials [34], the upper spinor component can also be derived from Eq. (18b) as

$$\begin{aligned}m\theta(x)&=\\&\left[E-\Lambda\frac{(1-e^{-2kx})}{(1+e^{-2kx})}-i\sqrt{m^2-(E-\Lambda)^2}-\frac{kie^{-2kx}}{(1+e^{-2kx})}\left(1+\frac{1}{k}\sqrt{k^2+4\left(ik\Lambda-\Lambda^2\right)}\right)\right]\varphi(x)\\&-B_ne^{-\sqrt{m^2-(E-\Lambda)^2}x}\left(1+e^{-2kx}\right)^{\left(1+\sqrt{k^2+4\left(ik\Lambda-\Lambda^2\right)}/k\right)/2}2kne^{-2kx}\\&\times\frac{\sqrt{m^2-(E-\Lambda)^2}/k+\sqrt{k^2+4\left(ik\Lambda-\Lambda^2\right)}/k+n+1}{\sqrt{m^2-(E-\Lambda)^2}/k}\end{aligned}$$

$$\times {}_2F_1\left(1-n,\sqrt{m^2-\left(E-\Lambda\right)^2}/k+\sqrt{k^2+4\left(ik\Lambda-\Lambda^2\right)}/k+n+2;\sqrt{m^2-\left(E-\Lambda\right)^2}/k+2;-e^{-2kx}\right). \quad (29)$$

Now, we calculate the normalized constant of the wave function satisfying the normalization condition

$$\int_0^1 \left|\varphi_n(s)\right|^2 ds = 1. \quad (30)$$

Two different forms of Jacobi polynomials [32, 34] are

$$P_n^{(c,d)}(x) = 2^{-n}\sum_{p=0}^{n}(-1)^{n-p}\binom{n+c}{p}\binom{n+d}{n-p}(1-x)^{n-p}(1+x)^p, \quad (31)$$

$$P_n^{(c,d)}(x) = \frac{\Gamma(n+c+1)}{n!\Gamma(n+c+d+1)}\sum_{r=0}^{n}\binom{n}{r}\frac{\Gamma(n+c++d+r+1)}{\Gamma(n+c+1)}\left(\frac{x-1}{2}\right)^r, \quad (32)$$

where

$$\binom{n}{r} = \frac{n!}{r!(n-r)!} = \frac{\Gamma(n+1)}{\Gamma(r+1)\Gamma(n-r+1)}. \quad (33)$$

By using the above equation, we have

$$P_n^{(2c,2d)}(1-2s) = (-1)^n\Gamma(n+2c+1)(n+2d+1) \times\sum_{p=0}^{n}\frac{(-1)^r}{p!(n-p)!}\frac{s^{n-p}(1-s)^p}{\Gamma(p+2d+1)\Gamma(n+2c-p+1)}, \quad (34)$$

$$P_n^{(2c,2d)}(1-2s) = (-1)^n\frac{\Gamma(n+2c+1)}{\Gamma(n+2c+2d+1)} \times\sum_{r=0}^{n}\frac{(-1)^r}{r!(n-r)!}\frac{\Gamma(n+2c+2d+r+1)}{\Gamma(2c+r+1)}s^r, \quad (35)$$

therefore, we have

$$1 = B_n^2(-1)^n\frac{\Gamma(n+2c+1)^2\Gamma(n+2d+1)}{\Gamma(2c+2d+1)} \times\sum_{p,r=0}^{n}\frac{(-1)^{p+r}\Gamma(n+2c+2d+r+1)I_n(p,r)}{p!r!(n-p)!(n-r)!\Gamma(p+2d+1)\Gamma(r+2d+1)\Gamma(n+2c-p+1)}, \quad (36)$$

where

$$I_n(p,r) = \int_0^1 s^{n+2\sqrt{\varepsilon}+r-p}(1-s)^{p-\mu P+1}ds, \quad c=\sqrt{\varepsilon}, \quad d=-\frac{1}{2}\mu P. \quad (37)$$

By using the following integral of hypergeometric function:

$$\int_0^1 s^{a-1}(1-s)^{c-a-b-1}ds = {}_2F_1(a,b;c;1)\frac{\Gamma(a)\Gamma(c-a)}{\Gamma(c)}, \quad (38)$$

which gives

$$\int_0^1 s^{a-1}(1-s)^{-b}ds = \frac{1}{a}\,{}_2F_1(a,b;a+1;1). \tag{39}$$

We obtain $I_n(p,r)$ from Eqs. (37) and (39) as

$$I_n(p,r) = \frac{1}{(n+2\sqrt{\varepsilon}+r-p+1)} \times {}_2F_1\left(n+2\sqrt{\varepsilon}+r-p+1, \mu P-p-1; n+2\sqrt{\varepsilon}+r-p+2;1\right). \tag{40}$$

**3.1.2. The case of $x<0$**

Now, we obtain the expression for $\varphi(x)$ in order to solve Eq. (20) for $x<0$, using the new variable $\tilde{s}=-e^{2kx}$, we have

$$\left\{\frac{d^2}{ds^2}+\frac{1-\tilde{s}}{\tilde{s}(1-\tilde{s})}\frac{d}{ds}+\frac{1}{\left[\tilde{s}(1-\tilde{s})\right]^2}\times \left[\frac{-\left(m^2-(E-\Lambda)^2\right)\tilde{s}^2+\left(2\Lambda^2-i4\Lambda k-2E^2+2m^2\right)\tilde{s}-\left(m^2-(E+\Lambda)^2\right)}{4k^2}\right]\right\}\varphi(\tilde{s})=0 \tag{41}$$

we obtain the coefficients

$$c_1=1, \qquad A=\frac{m^2-\left(E-\Lambda\right)^2}{4k^2},$$

$$c_2=1, \qquad B=\frac{(-4ik\Lambda+2\Lambda^2-2E^2+2m^2)}{4k^2},$$

$$c_3=1, \qquad C=\frac{m^2-\left(E+\Lambda\right)^2}{4k^2}. \tag{42}$$

In the case $x<0$, the specific values of coefficients $c_i$ $(i=4,5,...,9)$ are found from Eqs. (9) while $c_i$ $(i=10,11,12,13)$ from Eq. (14) as displayed in Table 1.

and using the NU method by following similar procedures presented in Subsect. 3.1.1., the energy eigenvalues equation for this case as

$$\sqrt{\frac{k^2+4\left(ik\Lambda-\Lambda^2\right)}{k^2}}=\sqrt{\frac{m^2-\left(E+\Lambda\right)^2}{k^2}}+\sqrt{\frac{m^2-\left(E-\Lambda\right)^2}{k^2}}-\left(2n+1\right). \tag{43}$$

On the other hand, the wave function can be calculated via Eq. (13) and Table 1 ($x<0$ case) as

$$\varphi(x)=Ne^{\sqrt{m^2-(E+\Lambda)^2}x}(1+e^{2kx})^{\frac{1}{2}\left(1+\frac{1}{k}\sqrt{k^2+4\left(ik\Lambda-\Lambda^2\right)}\right)}P_n^{\left(-\frac{1}{k}\sqrt{m^2-(E+\Lambda)^2},\frac{1}{k}\sqrt{k^2+4\left(ik\Lambda-\Lambda^2\right)}\right)}(1+2e^{2kx}), \quad (44)$$

or equivalently

$$\varphi(x)=N_n e^{\sqrt{m^2-(E+\Lambda)^2}x}(1+e^{2kx})^{\frac{1}{2}\left(1+\frac{1}{k}\sqrt{k^2+4\left(ik\Lambda-\Lambda^2\right)}\right)}$$
$$\times {}_2F_1\left(-n,-\sqrt{m^2-\left(E+\Lambda\right)^2}/k+\sqrt{k^2+4\left(ik\Lambda-\Lambda^2\right)}/k+n+1;-\sqrt{m^2-\left(E+\Lambda\right)^2}/k+1;\right.$$

.

where $x<0$ and where $N_n=N\dfrac{\Gamma\left(n-\sqrt{m^2-\left(E+\Lambda\right)^2}/k+1\right)}{n!\Gamma\left(-\sqrt{m^2-\left(E+\Lambda\right)^2}/k+1\right)}.$

is the normalized constant that is determined in similar procedures presented in Subsect. 3.1.1.

Following Eq. (41) of Ref. [33] and using the differential and recursion properties of the Jacobi polynomials [34], the upper spinor component can also be derived from Eq. (18b) as

$$m\theta(x)=$$
$$\left[E+\Lambda\frac{(1-e^{2kx})}{(1+e^{2kx})}+i\sqrt{m^2-\left(E+\Lambda\right)^2}+\frac{kie^{2kx}}{(1+e^{2kx})}\left(1+\frac{1}{k}\sqrt{k^2+4\left(ik\Lambda-\Lambda^2\right)}\right)\right]\varphi(x)$$
$$+N_n e^{\sqrt{m^2-(E+\Lambda)^2}x}\left(1+e^{2kx}\right)^{\left(1+\sqrt{k^2+4\left(ik\Lambda-\Lambda^2\right)}/k\right)/2}2kne^{2kx}$$
$$\times\frac{\sqrt{m^2-\left(E+\Lambda\right)^2}/k+\sqrt{k^2+4\left(ik\Lambda-\Lambda^2\right)}/k+n+1}{\sqrt{m^2-\left(E+\Lambda\right)^2}/k}$$
$$\times {}_2F_1\left(1-n,-\sqrt{m^2-\left(E+\Lambda\right)^2}/k+\sqrt{k^2+4\left(ik\Lambda-\Lambda^2\right)}/k+n+2;-\sqrt{m^2-\left(E+\Lambda\right)^2}/k+2;-e^{2kx}\right).. \quad (45)$$

**4. Approximate Solutions**

It is not straightforward to obtain an approximate expression for the energy eigenvalue of Eq. (29) and Eq. (44). It is also not possible to apply perturbation theory to find complex energy eigenvalues. In this section, we derive an approximate solution to Eq. (19) for kink-like potential at small values of $\Lambda$ and $k.$ The kink-like potential becomes the linear potential $\lambda x,$ where $\lambda=k\Lambda.$

The linear potential, $V(x)=\lambda x,$.is inserted in Eq. (19) giving

$$\frac{d^2\varphi(x)}{dx^2}+\left(\lambda^2x^2-2\lambda Ex-m^2+E^2+i\lambda\right)\varphi(x)=0. \quad (46)$$

Now we proceed to solve Eq. (46) demanding the solution to satisfy the resonance asymptotic conditions, i.e., we choose damping solutions for $x>0$ and diverging oscillating functions for $x<0$. We obtain approximate spectral equation of the form

$$2iE = 1+2n+\sqrt{1+4(m^2-E^2-ik\Lambda)},$$

which is energy eigenvalues for resonant states. The wave function are found as

$$\varphi(x) = Ax^{-n+iE}e^{-ik\Lambda x}L_n^{(i2E-2n-1)}(i2k\Lambda x), \quad x>0.$$

## 4. Conclusion

In this work, we have obtained the approximate energy eigenvalues equation and the corresponding normalized wave functions of the 1+1 dimensional Dirac equation in the presence of an attractive vectorial delta point interaction when one introduces perturbative potential of the form kink-like potential with small $\Lambda$ and $k$. We studied two cases when $x>0$ and $x<0$.and obtained the corresponding wave functions expressed in terms of the hypergeometric functions satisfying the boundary conditions. We used the powerful parametric generalization of the NU method in our solution.

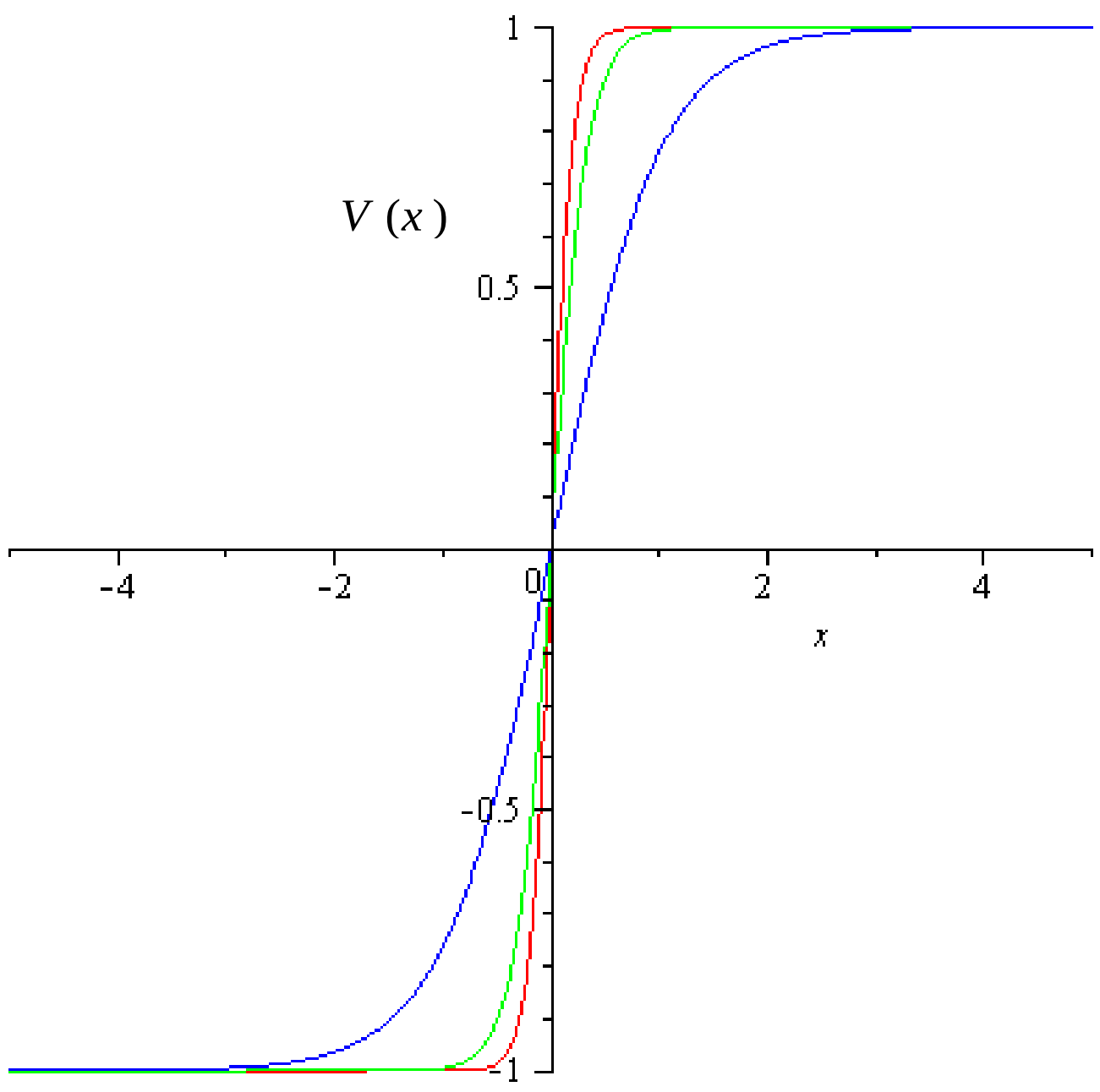

**Fig. 1** The variation of the hyperbolic potential as a function of $x$, for $\Lambda=1$ and three values of $k=1,3,5$.

**Table 1.** The specific values for the parametric constants necessary to calculate the energy eigenvalues and eigenfunctions in the case $x>0$.

| Constant | Analytic value for $x>0$: | Analytic value for $x<0$: |
|---|---|---|
| $c_4$ | 0 | 0 |
| $c_5$ | $-\frac{1}{2}$ | $-\frac{1}{2}$ |
| $c_6$ | $\frac{1}{4}+\frac{\left[m^2-(E+\Lambda)^2\right]}{4k^2}$ | $\frac{1}{4}+\frac{\left[m^2-(E-\Lambda)^2\right]}{4k^2}$ |
| $c_7$ | $\frac{(4ik\Lambda-2\Lambda^2+2E^2-2m^2)}{4k^2}$ | $\frac{(4ik\Lambda-2\Lambda^2+2E^2-2m^2)}{4k^2}$ |
| $c_8$ | $\frac{m^2-(E-\Lambda)^2}{4k^2}$ | $\frac{m^2-(E+\Lambda)^2}{4k^2}$ |
| $c_9$ | $\frac{1}{4}+\frac{ik\Lambda-\Lambda^2}{k^2}$ | $\frac{1}{4}+\frac{ik\Lambda-\Lambda^2}{k^2}$ |
| $c_{10}$ | $\frac{1}{k}\sqrt{m^2-(E-\Lambda)^2}$ | $-\frac{1}{k}\sqrt{m^2-(E+\Lambda)^2}$ |
| $c_{11}$ | $\sqrt{1+\frac{4\left(ik\Lambda-\Lambda^2\right)}{k^2}}$ | $\sqrt{1+\frac{4\left(ik\Lambda-\Lambda^2\right)}{k^2}}$ |
| $c_{12}$ | $\frac{1}{2k}\sqrt{m^2-(E-\Lambda)^2}$ | $-\frac{1}{2k}\sqrt{m^2-(E+\Lambda)^2}$ |
| $c_{13}$ | $\frac{1}{2}\left(1+\sqrt{1+\frac{4\left(ik\Lambda-\Lambda^2\right)}{k^2}}\right)$ | $\frac{1}{2}\left(1+\sqrt{1+\frac{4\left(ik\Lambda-\Lambda^2\right)}{k^2}}\right)$ |

## References

[1] Y.I., Salamin, S., Hu, K.Z., Hatsagortsyan, C.H., Keitel, Relativistic high-power laser–matter interactions, *Phys. Rep*. **427**(2–3) (2006) 41-155.

[2] M.I., Katsnelson, K.S., Novoselov, A.K., Geim, Chiral Tunnelling and the Klein Paradox in Graphene, *Nature Phys.* **2** (2006) 620-625.

[3] I.C., Wang, C.Y., Wong, Finite-size effect in the Schwinger particle-production mechanism, *Phys. Rev.* D **38** (1988) 348-359.

[4] C.-S. Jia, A. de Souza Dutra, Extension of PT-symmetric quantum mechanics to the Dirac theory with position-dependent mass, *Ann. Phys*. **323** (2008) 566-579.

[5] X.-L. Peng, J.-Y. Liu, C.-S. Jia, Approximation solution of the Dirac equation with position-dependent mass for the generalized Hulthen potential, *Phys. Lett*. A **352** (2006) 478-483.

[6] H. Akcay, Dirac equation with scalar and vector quadratic potentials and Coulomb-like tensor potential, *Phys. Lett*. A **373** (2009) 616-620.

[7] O. Aydogdu, R. Sever, Exact solution of the Dirac equation with the Mie-type potential under the pseudospin and spin symmetry limit, *Ann. Phys*. **325** (2010) 373-383.

[8] S. Zarrinkamar, A.A. Rajabi, H. Hassanabadi, Dirac equation for the harmonic Scalar and vector potentials and linear plus Coulomb-like tensor potential; the SUSY approach, *Ann. Phys*. **325** (2010) 2522-2528.

[9] M. Eshghi, H. Mehraban, Solution of the Dirac equation with position-dependent mass for q-parameter modified Poschl-Teller and Coulomb-like tensor potential, *Few-Body Syst*. **52** (2012) 41-47.

[10] M. Eshghi, M. Hamzavi, Spin symmetry in Dirac-Attractive Radial problem and tensor potential, *Commun. Theor. Phys*. **57** (2012) 355-360.

[11] M. Eshghi, Dirac-hyperbolic Scarf problem including a Coulomb-like tensor potential, *Acta Sci. Thech*. **34**(2) (2012) 207-215.

[12] M. Eshghi, H. Mehraban, Eigen spectra in Dirac-hyperbolic problem plus tensor coupling, *Chin. J. Phys*. Accepted and to appear in (2012).

[13] M. Eshghi, H. Mehraban, Eigen spectra for Manning-Rosen potential including a Coulomb-like tensor interaction, *Int. J. Phys. Sci*. **16** (2011) 6643-6652.

[14] M.V. Victor, A. G.-D. Luis, Particle resonance in the Dirac equation in the presence of a delta interaction and perturbative hyperbolic potential, *Europ. Phys. J. C* **61**(3) (2009) 519-525; *Arxiv*:0903.2597v2 [hep-th] 25 Mar 2009.

[15] A.G.-D. Luis, M.V. Victor, Resonances in the one-dimensional Dirac equation in the presence of a point interaction and a constant electric field, Phys. Lett. A 352 (2006) 202-205.

[16] W.-J. Tian, Bound state for spin-0 and spin-1/2 particles with vector and scalar hyperbolic tangent and cotangent potentials, http://www.paper.edu.cn.

[17] Antonio S. de Castro, M. Hott, Trapping neutral fermions with kink-like potentials, Phys. Lett. A **351** (2006) 379.

[18] Antonio S. de Castro, Effects of a mixed vector-scalar kink-like potential for spinless particles in two-dimensional space-time, Int. J. Mod. Phys. A 22 (2007) 2609; SOI:10.1142/S0217751X07036828.

[19] Chun-Sheng Jia, Yong-Feng Diao, Jian-Yi Liu, Bounded solutions of the Dirac equation with a PT-symmetric kink-like vector potential in two-dimensional space-time, Int. J. Theor. Phys. **47** (2008) 2513-2522.

[20] Chun-Sheng Jia, A. de Souza Dutra, Extension of PT symmetric quantum mechanics to Dirac theory with position-dependent mass, Ann. Phys. (N.Y.) **323** (2008) 566-579.

[21] Chun-Sheng Jia, Xiao-Ping Li, Lie-Hui Zhang, Exact solutions of the Klein-Gordon equation with position-dependent mass for mixed vector and scalar kink-like potentials, Few Body Syst. **52** (2012) 11-18.

[22] Victor M. Villalba, Luis A. González-Díaz, Particle resonance in the Dirac equation in the presence of a delta interaction and a perturbative hyperbolic potential, Eur. Phys. J. C **61** (3) (2009) 519-525; arXiv:0903.2597 .

[23] A.F. Nikiforov, V.B. Uvarov, *Special functions of mathematical physics*, Birkhauser, Verlag, Basel, 1988.

[24] C. Tezcan, R. Sever, A General Approach for the Exact Solution of the Schrödinger Equation, *Int. J. Theor. Phys*. **48** (2009) 337-350.

[25] S.M. Ikhdair, Rotational and vibrational diatomic molecule in the Klein-Gordon equation with hyperbolic scalar and vector potentials, Int. J. Mod. Phys. C **20** (10) (2009) 1563-1582.

[26] S.M. Ikhdair, Rotation and vibration of diatomic molecule in the spatially-dependent mass Schrödinger equation with generalized $q$-deformed Morse potential, Chem. Phys. **361** (2009) 9-17.

[27] W. Greiner, *Relativistic Quantum Mechanics, Wave Equations*, Springer Verlog, New-York, 1990.

[28] L.H. Ryder, *Quantum field theory*, Cambridge: Cambridge University Press, 1985.

[29] C.-S. Jia, A. de Souza Dutra, Position-dependent effective mass Dirac equations with *PT*-symmetric and non-*PT*-symmetric potentials, *J. Phys. A: Math. Gen*. **39**. (2006) 11877. doi:10.1088/0305-4470/39/38/013

[30] E.C. Titchmarsh, *Eigenfunction expansions associated with second order differential equations*, part II. Oxford University, 1958.

[31] W.E. Brittin, *Lectures in theoretical physics*, Vol. IV, Interscience Publishers, New-York, 1962.

[32] M. Abramowitz, I. Stegun, Handbook of mathematical function with formulas, graphs and mathematical tables, Dover, New York, 1964.

[33] S.M. Ikhdair, Approximate solutions of the Dirac equation for the Rosen-Morse potential including the spin-orbit centrifugal term, J. Math. Phys. **51** (2010) 023525.

[34] W. Magnus, F. Oberhenttinger, R.P. Soni, Formulas and theorems for the special functions of mathematical physics, 3Ed., Springer, Berlin, 1966.